\documentclass[pre,aps,floatfix,twocolumn,preprintnumbers,superscriptaddress,notitlepage,longbibliography]{revtex4-1}
\usepackage[utf8]{inputenc}
\usepackage[T1]{fontenc}

\usepackage{graphicx}
\usepackage{dcolumn}
\usepackage{bm}
\usepackage{amssymb}
\usepackage{pifont}
\usepackage{amsmath}
\usepackage{times}

\usepackage{epstopdf}
\usepackage{latexsym}
\usepackage{keyval}
\usepackage{ifthen}
\usepackage{moreverb}
\usepackage{color}
\usepackage[normalem]{ulem}

\usepackage[hidelinks,unicode=true]{hyperref}

\usepackage{lipsum}
\usepackage{autonum}

\def \dd {{\rm d}}

\newcommand{\fletxa}{\xrightarrow[\scriptscriptstyle \nu\to \infty]{}}

\graphicspath{{../3_figures/}}

\begin{document}
\title{Exact low-dimensional description for fast neural oscillations with low firing rates}
\author{Pau Clusella}
\email{pau.clusella@upc.edu}
\affiliation{Departament de Matemàtiques, Universitat Politècnica de Catalunya, 08242 Manresa, Spain}
\author{Ernest Montbrió}
\affiliation{Neuronal Dynamics Group, Department of Information and Communication Technologies,
Universitat Pompeu Fabra, 08018 Barcelona, Spain.}

\begin{abstract}
Recently, low-dimensional models of neuronal activity have been exactly derived  
for large networks of deterministic, Quadratic Integrate-and-Fire (QIF) neurons. Such
\emph{firing rate models} (FRM) describe the emergence of fast collective oscillations (>30~Hz)
via the frequency-locking of a subset of neurons to the global oscillation frequency.  
However, the suitability of such models to describe realistic neuronal states is 
seriously challenged by fact that during episodes of fast collective oscillations, 
neuronal discharges are often very irregular and have low firing rates
compared to the global oscillation frequency.
Here we extend the theory to derive exact FRM for QIF neurons to include noise, 
and show that networks of stochastic neurons displaying
irregular discharges at low firing rates during episodes of fast oscillations, are governed 
by exactly the same evolution equations as deterministic networks. 
Our results reconcile two traditionally confronted views on neuronal synchronization, 
and upgrade the applicability of exact FRM
to describe a broad range of biologically realistic neuronal states.
\end{abstract}

\maketitle

\section{Introduction}

Fast oscillations (>30~Hz) are a prominent feature of neural activity
~\cite{Bartos2007,Whittington1995,Whittington2000,Wan10}.
Empirical studies show that very often such rhythms display a remarkable dichotomy:
At the collective level, neuronal oscillations are fast and fairly regular,
whereas at the single-cell level
individual spikes trains remain highly irregular and have low firing 
rates~\cite{Wan10,BW12,Brunel2008}.

A wealth of theoretical and computational work have investigated the emergence of
fast neuronal rhythms, and identified minimal neurophysiological ingredients
that robustly produce them in large ensembles of spiking model neurons
~\cite{Whittington1995,Whittington2000,Wan10,BW12,Brunel2008,Boe17}.
According to these studies, fast oscillations emerge in populations of
inhibitory neurons with synaptic time constants and/or fixed delays, and sufficient drive to induce spiking.
Notably, in this idealized modeling framework of neuronal synchrony,
networks of spiking neurons with random connectivity and additive noise 
may display irregular spike discharges 
at low firing rates, akin to experimental observations~\cite{Brunel1999,Brunel2000}.
This so-called \emph{sparse synchronization}~\cite{Brunel2008} state, is also encountered in
all-to-all coupled networks with both multiplicative and additive noise~\cite{Brunel1999,Brunel2000},
or just with additive noise~\cite{Tiesinga2000,Brunel2006}.

An alternative and powerful tool to investigate fast neuronal
oscillations is to use reduced or simplified models 
---called neural-mass, or firing-rate models (FRM)---,
which describe the mean activity in a neuronal population
~\cite{Wilson1972,LopesdaSilva1974,Ahn1974,Freeman1975}.
Such FRM consist of one, or a few ordinary differential equations, 
and allow for a thorough understanding of the system's dynamics. 
Specifically, for the case of an inhibitory network with synaptic delays,
FRM exhibit fast oscillations similar to those observed
in numerical simulations of large networks of spiking neurons
~\cite{RBH05,RM11,Devalle2017,Devalle2018}.

Fast oscillations in FRM are often  
associated to the presence of sparse synchronization~\cite{Brunel2003,Brunel2008}. 
However, the fact that FRM are heuristic and 
not exactly obtained from a given network of spiking neurons,
impedes one from unambiguously linking the collective dynamics described by the FRM 
with that of individual neurons in the network.
Yet, a notable exception is a recently developed mean-field 
theory that allows for a proper mathematical derivation of the 
FRM corresponding to an all-to-all coupled ensemble of heterogeneous,  
Quadratic Integrate-and-Fire (QIF) neurons~\cite{Montbrio2015,Luke2013}. 
Accordingly, this theory is singularly suited to 
investigate the relation between microscopic, single-cell dynamics (observed in 
numerical simulations of networks of spiking neurons), with that of the network's 
collective states ---exactly described by the QIF-FRM.

Unfortunately, the theory to obtain QIF-FRM
(also known as `next-generation neuron mass models'~\cite{Coombes2019}),
is only valid for deterministic networks with Cauchy heterogeneity, which 
are not capable of displaying sparsely synchronized states.
Indeed, synchronization emerging in populations of heterogeneous inhibitory neurons 
is due to the frequency entraintment of a subset of neurons, which 
display regular, periodic dynamics with the (fast) frequency of the collective rhythm
~\cite{Wang1996,White1998,Tiesinga2000,Devalle2017,Coombes2019,Laing2015,Pazo2016,RP18,Devalle2018,Bi2020,Segneri2020,Ceni2020,KBF+19,PP22,Clusella2022,PP23}.
This synchronization scenario for deterministic neurons is in sharp contrast with 
the sparse synchronization scenario, and seriously challenges the suitability of QIF-FRM to 
describe and investigate biologically plausible neuronal states. In addition, 
synchronization in such heterogeneous networks is considered to be fragile and at 
odds with sparse synchronization~\cite{Wan10,BW12,Brunel2008,Tiesinga2000}. 

Motivated by recent advances in the context of the Kuramoto model~\cite{Tanaka2020,Tonjes2020},
in this article we extend the theory to derive QIF-FRM to networks 
of QIF neurons driven by Cauchy noise. Strikingly, the resulting QIF-FRM 
reveals that deterministic QIF networks showing 
fast oscillations via frequency-entraintment are governed by the same evolution 
equations as networks of stochastic QIF neurons displaying sparse synchronization. 

\section{Synchronization scenarios in populations of inhibitory QIF neurons}

\begin{figure*}[t]
  \centerline{\includegraphics[width=\textwidth]{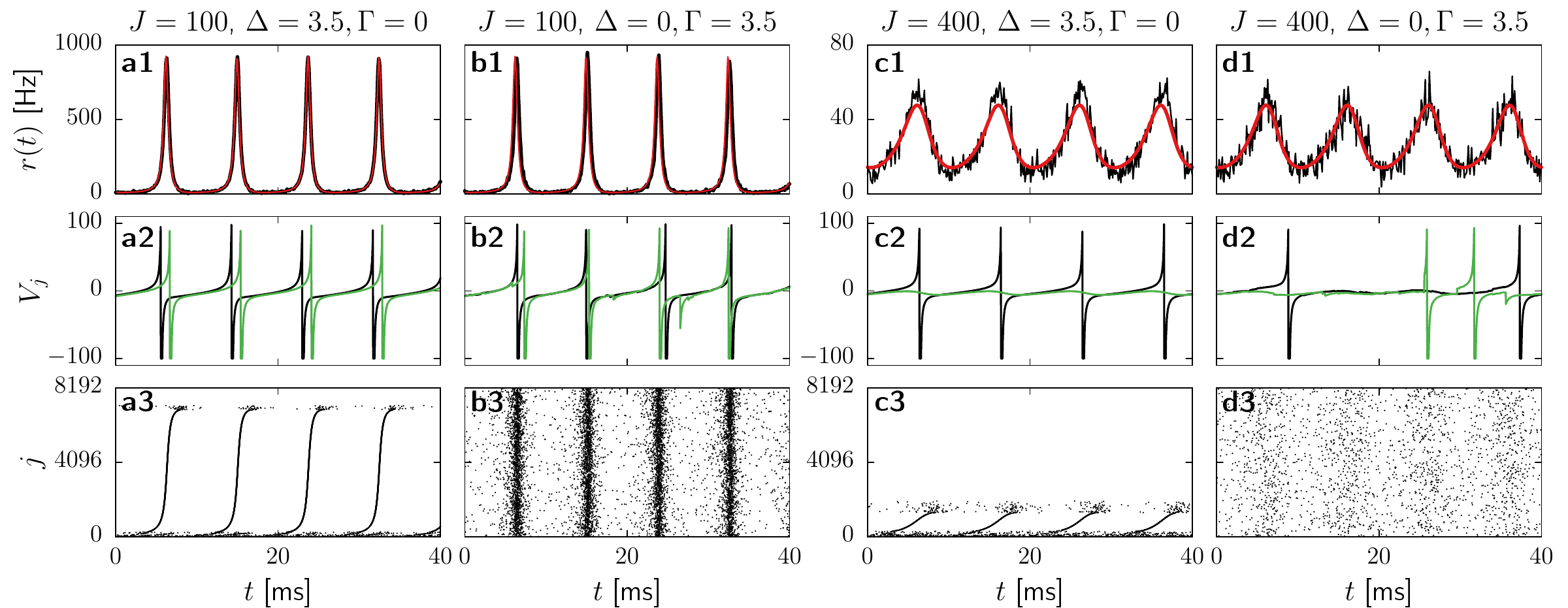}}
\caption{
Synchronization scenarios in an inhibitory network of globally coupled QIF neurons,
Eqs.~(\ref{eq:qif},\ref{eq:synapses}),
with quenched Cauchy heterogeneity (panels (a1)--(a3) and (c1)--(c3)), and with Cauchy noise (panels (b1)--(b3) and (d1)--(d3)),
and $\overline \eta=100$.
Panels (a1)--(d1): Black lines: Time series of the mean firing rate
Eqs.~(\ref{eq:qif},\ref{eq:synapses}).
Red lines: Time series of the $r$-variable of the QIF-FRM~(\ref{eq:fre},\ref{eq:synapses}).
Panels (a2)--(d2): Time series of the membrane potential of two individual QIF neurons.
Panels (a3)--(d3): Raster plots of the spiking times of a subset of $25\%$ of randomly selected neurons.
Neuron indices in panels (a3) and (c3) are sorted according to $\eta_j$.
}
  \label{figure1}
\end{figure*}

We consider a population of $N$ QIF
neurons~\cite{EK86,Izhikevich2007}, interacting via a mean-field inhibitory 
coupling of strength $J$. 
The evolution of the membrane potential of a QIF neuron obeys the equation   
\begin{equation}\label{eq:qif}
\tau_m \dot V_j=V_j^2 +\eta_j  + \xi_j(t) -\tau_m Js(t),
\end{equation}
where $j=1,\dots,N$, and  the following resetting rule: if $V_j>V_p$ then $V_j\leftarrow V_r$.
The neuron's membrane time constant $\tau_m$ is set to 10~ms, and  
quenched heterogeneity is modeled via   
parameter $\eta_j$, which represents a constant input current that varies from neuron to neuron 
according to a Cauchy probability density function $G(\eta)$,
centered at $\overline \eta$ and with half-width at half-maximum (HWHM) $\Delta$,
\begin{equation}
G(\eta):=\frac{1}{\pi}\frac{\Delta }{(\eta-\overline \eta)^2+\Delta^2}\;.
\end{equation}
In addition, neurons are subject to independent noisy inputs. Specifically, the random variables 
$\xi_j(t)$ represent zero-centered Cauchy white noise with HWHM $\Gamma$.
Finally, neurons interact all-to-all through the 
mean post-synaptic activity $s(t)$. This mean field variable is related via  the equation
\begin{equation}\label{eq:synapses}
\tau_s \dot s(t)=-s(t)+r(t),
\end{equation}
to the population mean firing rate $r(t)$
\begin{equation}\label{eq:selfconsistent2}
r(t)=\frac{1}{N}\sum_{j=1}^N\sum_k\frac{1}{\tau_r}\int_{t-\tau_r}^t\dd \zeta\; \delta(\zeta-t_j^{(k)})\;.
\end{equation}
The time constant $\tau_s$ in Eq.~\eqref{eq:synapses} corresponds to the synaptic decay time 
of the inhibitory synapses, which we set to $\tau_s=5$~ms.  
The instant $t_j^{(k)}$ in Eq.\eqref{eq:selfconsistent2} indicates the $k$-th spike of neuron $j$, 
and $\tau_r$ is a time window of the spike events,
which we set to $10^{-2}$~ms
~\footnote{The mean firing rate in Eq.~\eqref{eq:selfconsistent2} exactly corresponds 
to the firing rate variable described by Eqs.~\eqref{eq:fre}
if one first adopts the limit $N\to\infty$, and then $\tau_r\to 0$, see~\cite{Montbrio2015}.
}.

In Figure~\ref{figure1}, we compare the results of numerical simulations
\footnote{In numerical simulations of Eqs.~(\ref{eq:qif},\ref{eq:synapses}),
we used the Euler-Maruyama scheme with $\delta t=10^{-3}$.
Thus, at each time step, the random increments follow Cauchy distribution with
HWHM $dt \Gamma$.
In addition, we set $V_p=-V_r=100$, $\overline \eta=100$ and $N=8192$.
For heterogenous networks, $\eta_j$ are generated
as $\eta_j=\Delta \tan(\frac{\pi(2j-N-1)}{2(N+1)})$, with $j=1, \dots N$.
The mean firing rate $r$ is computed using Eq.~\eqref{eq:selfconsistent2} with
$\tau_r=10^{-2}$. The time series of Fig~\ref{figure1}(a1--d1)
use an additional binning window of $0.1$ms for display.}
in two different synchronous regimes of Eqs.~(\ref{eq:qif},\ref{eq:synapses}), using
deterministic and noisy networks.
For moderated inhibition, the mean firing rate $r(t)$
displays fast oscillations at approximately $100$~Hz, which are noticeably similar in
both the
deterministic and the noisy network, see Fig.~\ref{figure1}(a1) and Fig.\ref{figure1}(b1), respectively.
The large amplitude of the firing rate oscillations reflects a high degree of synchronization.
Indeed, Figs.~\ref{figure1}(a2) and \ref{figure1}(b2) show the membrane potential of two neurons
in the heterogeneous and noisy networks respectively.
In both cases, neurons
fire periodically with the frequency of the global oscillations.
Additionally, the raster plots in Figs.~\ref{figure1}(a3) and ~\ref{figure1}(b3)
confirm that most neurons display
such regular, periodic dynamics.

In contrast, for strong inhibitory coupling ($J=400$),  the amplitude of the
oscillations is greatly reduced (see Figs.~\ref{figure1}(c1,d1), and the
striking similarity between the firing rate dynamics of the deterministic and the stochastic networks)
and the microscopic states of the two networks strongly differ.
Indeed,
the raster plot of the deterministic population Fig~\ref{figure1}(c3) shows that only a
small subset of the neurons fire regularly, while a majority is
strongly suppressed due to feedback inhibition~\cite{White1998}.
On the other hand, in the stochastic network, noise may release some of the neurons from
suppression, producing highly irregular spike trains, with low firing rates and little
indication of the collective oscillation, see Figs.~\ref{figure1}(d2,d3).
This latter state corresponds to the sparse synchronization originally uncovered in networks
of inhibitory neurons~\cite{Brunel1999,Brunel2000}.

\begin{figure}[t]
  \centerline{\includegraphics[width=0.45\textwidth]{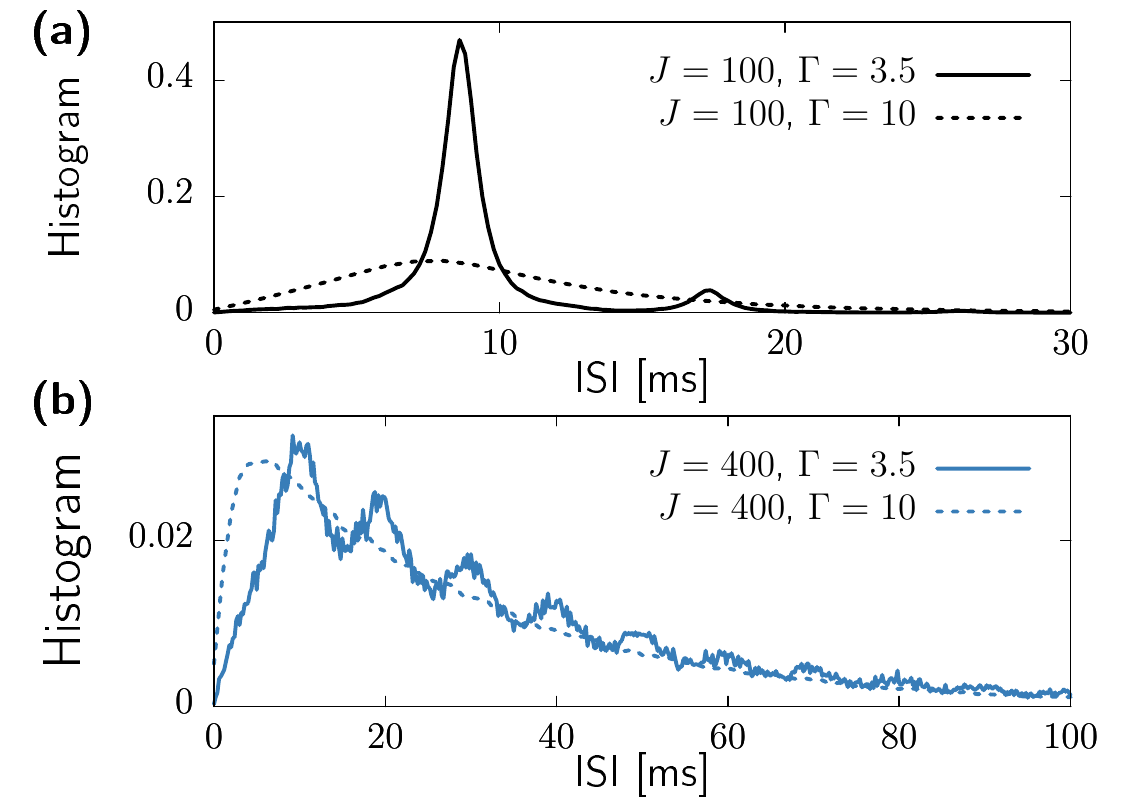}}
        \caption{
                Interspike interval (ISI) histograms of populations of stochastic QIF neurons for $J=100$
                (panel a) and $J=400$ (panel b).
                Continuous lines correspond to the results in Fig~\ref{figure1}(b3,d3) ($\Gamma=3.5$),
                dashed lines correspond to a asynchronous state  ($\Gamma=10$).
   }
  \label{figureISI}
\end{figure}

To further emphasize the effects of noise in synchronized states, in Fig.~\ref{figureISI} (solid lines)
we computed the distribution of interspike intervals (ISI) corresponding to the
synchronization regimes shown in columns b and d of Fig.~\ref{figure1}.
Figure~\ref{figureISI}(a) indicates that, for weak inhibition, the distribution
has a large peak that coincides with the collective oscillation period
($T\approx 8.7$ms). In addition, there is a small resonance in a second harmonic,
indicating that most neurons fire once per cycle and, due to noise, some skipping events occur.
The corresponding Coefficient of Variation ---quantifying the broadness of the ISI distribution---
is $\text{CV}\approx 0.35$,
which confirms that, despite the presence of noise, spike trains remain highly regular
\footnote{The coefficient of variation (CV) of the ISI provides a common measure of firing regularity.
It is close to 0 for a delta distribution,
close to 1 for Poisson spike times,
and larger for even more irregular patterns.}.
Figure~\ref{figureISI}(b) shows the ISI histogram for strong inhibition.
In this case the distribution is broad ($\text{CV}=0.85$),
spanning several periods of the oscillation cycle,
with small peaks located at the harmonics of the fundamental period.
This indicates that spike trains are highly irregular, and close to a Poisson process ($\text{CV}=1$).
In asynchronous regimes, the additional peaks of the ISI distribution vanish, 
leading to a unimodal histogram (see dashed lines in Fig.~\ref{figureISI}(a,b)).

\section{Exact Firing Rate Model}

In the thermodynamic limit, Eqs.~(\ref{eq:qif},\ref{eq:synapses}) with $\Gamma=0$
are exactly described by a low-dimensional system of differential equations that  
we refer to as QIF-FRM~\cite{Montbrio2015}.
Inspired by recent results in the context of the Kuramoto model~\cite{Tonjes2020,Tanaka2020},
we next show that an identical set of exact FRM is obtained if, in addition to
heterogeneity, neurons are driven by independent Cauchy noise 
\footnote{
See Ref.~\cite{Pietras2023} for an alternative derivation.
Additionally, see also recent attempts to obtain \emph{approximated} mean field
descriptions of populations of QIF neurons with independent Gaussian noise
~\cite{Goldobin2021,RP19,Goldobin2021_2,DiVolo22}.}.\\

\subsection{Fractional Fokker-Planck equation and Lorentzian ansatz}

Adopting the thermodynamic limit, $N\to\infty$, 
the macroscopic state of the QIF network is given by the probability density function of neurons 
having membrane potential $V$ at time $t$,
\begin{equation}\label{eq:Q}
	Q(V,t)=\int_{-\infty}^\infty P(V,t;\eta)G(\eta)\dd \eta\;,
\end{equation}
where  $P(V,t;\eta)$ are conditional densities for specific values of $\eta$.
%
In the thermodynamic limit, $N\to\infty$,
the time evolution of the  conditional
densities $P(V,t;\eta)$ is given by a
Fractional Fokker-Planck Equation (FFPE)~\cite{metzler2000,klafter2011,Mendez2014}.
This equation involves the Riesz fractional derivative, which is usually defined in Fourier space (for an equivalent description that avoids the use of Riesz derivatives see Appendix~\ref{appA}).
We introduce the Fourier transform and its inverse as
\begin{equation}
        \mathcal{F}\{f(V,t)\}=\int_{-\infty}^{\infty} f(V,t)e^{i k V} \dd V
\end{equation}
and
\begin{equation}
        \mathcal{F}^{-1}\{\tilde f(k,t)\}=\frac{1}{2\pi}\int_{-\infty}^{\infty} \tilde f(k,t)e^{-ikV}\dd k\;.
\end{equation}
Then, the Riesz operator $\frac{\partial^{\alpha}}{\partial |V|^\alpha} P(V,t)$ is defined as
\begin{equation}\label{eq:riesz}
        \mathcal{F}\left\{\frac{\partial^\alpha f(V,t)}{\partial |V|^{\alpha}}\right\}=-|k|^{\alpha} \mathcal{F}\{f(V,t)\}\;.
\end{equation}
For Cauchy noise we are only interested in the case $\alpha=1$.
With this definition, the FFPE for the time evolution of the densities $P$ reads
\begin{align}\label{eq:ffpe}
	&\tau_m\frac{\partial P}{\partial t}(V,t;\eta) =\\
	&-\frac{\partial}{\partial V}\left\{\left[ V^2+\eta-\tau_m Js(t)\right] P(V,t;\eta)\right\}
	+\Gamma \frac{\partial P}{\partial |V|}(V,t;\eta) \;.
\end{align}
For $\Gamma=0$, we recover the continuity equation for deterministic dynamics. 
This case was solved in~\cite{Montbrio2015} assuming that the conditional probabilities
$P$ are Cauchy distributions 
with width and center parameters that depend on $t$ and $\eta$, namely $x(t,\eta)$ and $y(t,\eta)$,
\begin{equation}\label{eq:ansatz}
        P(V,t,\eta)=\frac{1}{\pi}\frac{x(t,\eta)}{[V-y(t,\eta)]^2+x(t,\eta)^2}\;.
\end{equation}
Here, we employ the same Lorentzian ansatz to solve the FFPE~\eqref{eq:ffpe} for arbitrary $\Gamma$.

The Fourier transform of $P$ reads $\tilde P(k,t)=\exp\{iky-|k|x\}$, thus,
\begin{align}
        \frac{\partial P}{\partial |V|}(V,t)
        &=\frac{-1}{2\pi}\int_{-\infty}^\infty |k|e^{-ik(V-y)-|k|x}\dd k\\
        &=\frac{1}{\pi}\frac{(V-y)^2 - x^2}{[(V-y)^2 + x^2]^2},\label{eq:equiv}
\end{align}
where we have performed the integrals by parts.
We replace this last expression (Eq.~\eqref{eq:equiv}) in the FFPE~\eqref{eq:ffpe}, and expand the partial derivatives using
Eq.~\eqref{eq:ansatz}.
After simplifying, we obtain a polynomial equation for $V$, which can be solved
by equating the coefficients of like powers of $V$ on both sides of the expression.
As a result, we obtain two differential equations for
the time evolution of $x(t,\eta)$ and $y(t,\eta)$:
\begin{equation}
\begin{aligned}\label{eq:xy}
        \tau_m \dot{x}(t,\eta) &= \Gamma + 2x(t,\eta)y(t,\eta) \\
        \tau_m \dot{y}(t,\eta) &= \eta+y(t,\eta)^2-x(t,\eta)^2 - J\tau_m s(t)\;.
\end{aligned}
\end{equation}

In a homogeneous population of neurons ($\Delta=0$) these two equations would already
provide a macroscopic description of the system.
For a heterogeneous population with Cauchy-distributed heterogeneities ($\Delta>0$), a low-dimensional system can be attained
by solving the integral in Eq.~\eqref{eq:Q}.
In order to do so we define the complex variables $w(t,\eta):= x(t,\eta)+i y(t,\eta)$, thus
\begin{equation}\label{eq:w}
        \tau_m \dot w(t,\eta) = i[ \eta - w(t,\eta)^2 + J \tau_m s(t)]+\Gamma\;.
\end{equation}
Considering the analytic continuation of $w(t,\cdot)$
to the complex plane provides  $P(V,t;\cdot)$ as a holomorphic function.
Therefore, we can compute the integral in Eq.~\eqref{eq:Q} using Cauchy's residue theorem
along the closed semicircumference $|\eta| e^{i\theta}$ with $\theta \in (-\pi,0)$ and $|\eta|\to\infty$~\cite{Montbrio2015}.
As a result we obtain
\begin{equation}\label{eq:sol}
        Q(V,t)
        = P(V,t;\overline \eta - \Delta i) \;.
\end{equation}

\subsection{Mean membrane potential and mean firing rate}

Equation~\eqref{eq:sol} shows that $Q$ is a Lorentzian distribution.
Its center corresponds to the mean membrane potential of the QIF population, and thus we denote it as $v(t):=y(t,\overline \eta -\Delta i)$
\footnote{Notice that Lorentzian distributions do not have a well defined mean, thus $v$ actually corresponds to the Cauchy principal value of the integral $\int_{-\infty}^{\infty} Q(V,t)V\dd V$.}.
On the other hand, the mean firing rate of the neural population is given by
\begin{equation}\label{eq:integralrate}
        r(t)=\int_{-\infty}^{\infty} G(\eta)\tilde r(t,\eta)\dd \eta
\end{equation}
where $\tilde r(t,\eta)$ is the firing rate of the subset of neurons with current $\eta$.
This quantity can be computed as the probability flux of the FFPE~\eqref{eq:ffpe} at $V\to\infty$.
The probability flux of~\eqref{eq:ffpe} at a given point $V$ is given by two terms:
First, the flux given by the deterministic flow of the QIF dynamics,
\begin{equation}
        \tau_m^{-1}\left[ V^2+\eta+\tau_m Js(t)\right] P(V,t;\eta).
\end{equation}
Second, the total probability change rate provided by the stochastic dynamics in the interval $(V,\infty)$.
Using Eq.~\eqref{eq:equiv} this can be computed as
\begin{equation}
        \frac{\Gamma}{\pi\tau_m} \int_{V}^{\infty}  \frac{(U-y)^2 - x^2}{[(U-y)^2 + x^2]^2}\dd U = \frac{\Gamma}{\pi \tau_m} \frac{V-y}{x^2+(V-y)^2}\;.
\end{equation}
Altogether, we have that
\begin{align}\label{eq:firingrates}
	\tilde r(t,\eta) = \lim_{V\to\infty} \frac{1}{\tau_m}&\Bigl\{\left[ V^2+\eta+\tau_m Js(t)\right] P(V,t;\eta)\\
	&+\frac{\Gamma}{\pi \tau_m}\frac{V-y}{x^2+(V-y)^2}\Bigr\} = \frac{x(t,\eta)}{\tau_m \pi}\;.
\end{align}
Replacing this expression in Eq.~\eqref{eq:integralrate}
provides the firing rate of the entire QIF population as $r(t)=x(t,\overline \eta - \Delta i)/(\pi\tau_m)$.\\

Finally, replacing $y=v$, $x=\pi \tau_m r$, and $\eta=\overline \eta - \Delta i$ in Eq~\eqref{eq:w} and taking real and imaginary part
 leads to
\begin{equation}
\begin{aligned}\label{eq:fre}
        \tau_m \dot r &= \frac{\Gamma + \Delta}{\pi \tau_m } + 2rv \\
        \tau_m \dot v &= \overline \eta  + v^2 - (\pi\tau_m r)^2 - J\tau_m s \;,
\end{aligned}
\end{equation}
which, together with Eq.~\eqref{eq:synapses},
exactly describe the behavior of the QIF network Eqs.~(\ref{eq:qif},\ref{eq:synapses}).
Remarkably, Eqs.~(\ref{eq:fre}) illustrate that, in the thermodynamic limit,  the level of
heterogeneity $\Delta$ and the level of noise $\Gamma$ play identical roles
at the collective level in populations of globally coupled QIF neurons.

\section{The QIF-FRM captures fast oscillations with low firing rates}

For the deterministic case, $\Gamma=0$, the dynamics of Eqs.~~(\ref{eq:fre},\ref{eq:synapses}) 
have been analyzed in~\cite{Devalle2017}. In the following 
we extend this analysis to networks of QIF neurons with both heterogeneity and noise.

\begin{figure}[t]
  \centerline{\includegraphics[width=0.5\textwidth]{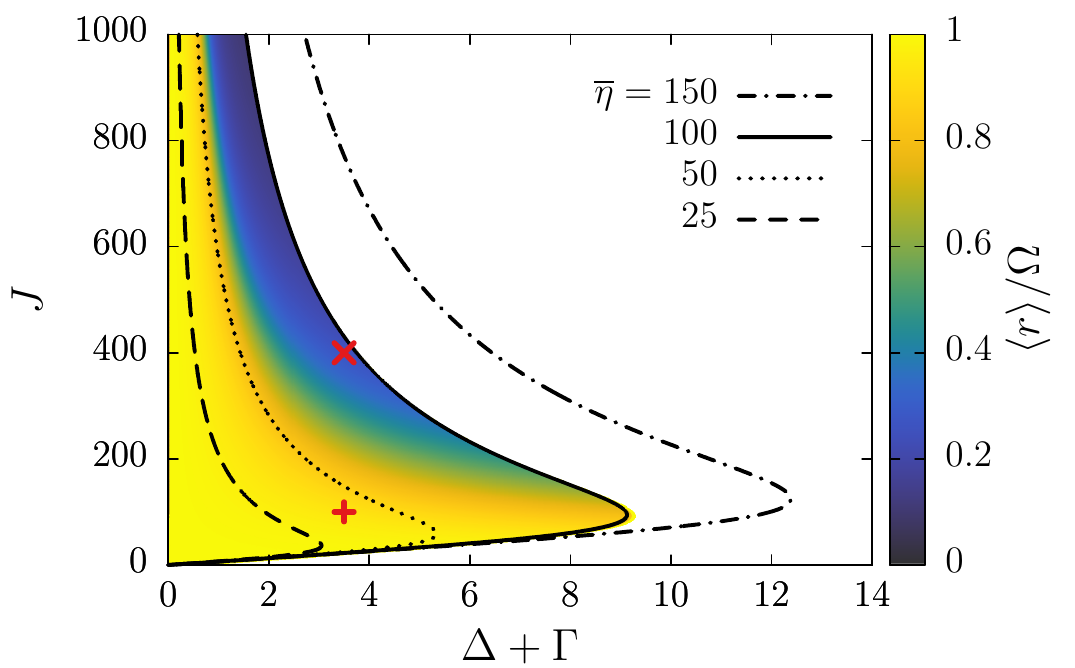}}
        \caption{Phase diagram of Eqs.~(\ref{eq:fre},\ref{eq:synapses}).
        Black lines: Supercritical Hopf bifurcations 
        for different values of $\overline \eta$, 
	obtained using AUTO-07p~\cite{auto07p}.
        The colormap corresponds to the average number of spikes per oscillation cycle for 
	$\overline \eta=100$,  and numerically obtained 
	computing $\Omega$ and $\langle r \rangle$ using the QIF-FRM,
	Eqs.~(\ref{eq:fre},\ref{eq:synapses}).
        Symbol $+$: parameter values corresponding to Fig.~\ref{figure1} panels (a1)--(a3) and (b1)--(b3);\
        Symbol $\times$: parameter values corresponding to Fig.~\ref{figure1} panels (c1)--(c3) and (d1)--(d3).
        }
  \label{figurePD}
\end{figure}

Eqs.~(\ref{eq:fre},\ref{eq:synapses}) have a single fixed-point corresponding to an 
asynchronous state. For small enough disorder $\Delta+\Gamma$ and $\tau_s,\overline \eta>0$,
this steady state loses stability via a supercritical
Hopf bifurcation, leading to the emergence of fast oscillations.
The phase diagram Fig.~\ref{figurePD} shows the Hopf boundaries
for different values of $\overline \eta>0$
as a function of the strength of inhibition, $J$, and the level of disorder, $\Delta+\Gamma$. 
Oscillations occur for small enough values of the disorder, 
i.e. to the left of the the Hopf boundaries.

To characterize neuronal activity within the region of oscillations, we compare  
the frequency of the collective rhythm $\Omega$, 
with the mean firing frequency of the 
individual neurons, given by the time-averaged mean firing rate $\langle r \rangle$.
Notably, the ratio $\langle r\rangle/\Omega$ (measuring the average spiking activity per 
oscillation cycle),  is independent of whether the network is heterogeneous, stochastic, or both heterogeneous and 
stochastic ---see the colored region in Fig.~\ref{figurePD}, which shows $\langle r\rangle/\Omega$ 
for $\overline \eta=100$ computed from numerical simulations of the FRM~(\ref{eq:fre},\ref{eq:synapses}).

For moderate disorder, Fig.~\ref{figurePD} shows that the transition from asynchronous to 
synchronous activity occurs in two different ways depending on $J$. 
For small inhibition most neurons behave as self-sustained oscillators 
frequency entrained by the collective oscillation.  
This case corresponds to the yellow regions in Fig.~\ref{figurePD}, and to   
columns (a) and (b) of Fig.~\ref{figure1}. 

By contrast, for strong $J$ suppression of firing dominates, and oscillations   
are only maintained by a few active neurons, see blue region in Fig.~\ref{figurePD}. 
In this case the population firing rate becomes considerably smaller than the oscillation frequency,
see columns c and d of Figs~\ref{figure1}.
For the case of stochastic neurons, this corresponds to sparse synchronization.

\begin{figure}[t]
  \centerline{\includegraphics[width=0.45\textwidth]{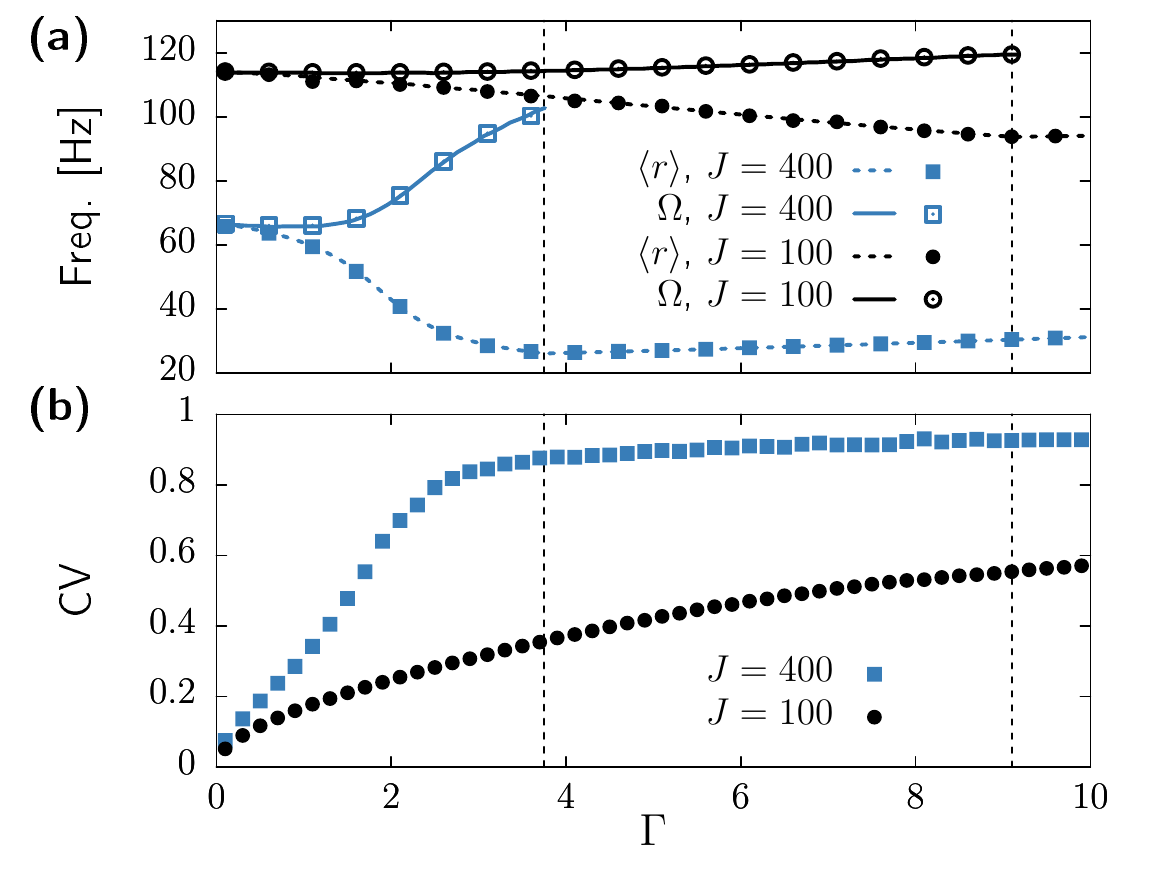}}
        \caption{(a) Oscillation frequency $\Omega$ and time-averaged firing rate $\langle r \rangle$ 
        of a homogeneous network ($\Delta =0$) of QIF neurons vs. noise intensity $\Gamma$.
	Results obtained using Eqs.~(\ref{eq:qif},\ref{eq:synapses}) (symbols) and 
	Eqs.~(\ref{eq:fre},\ref{eq:synapses}) (lines),
        for $J=100$ (black) and $J=400$ (blue).
        Vertical dashed lines indicate the location of the Hopf bifurcations 
        for $J=100$ ($\Gamma\simeq 9.11$) and for $J=400$ ($\Gamma\simeq 3.75$).
        (b) Coefficient of variation (CV) of the ISI, obtained using the network 
        Eqs.~(\ref{eq:qif},\ref{eq:synapses}). 
   }
  \label{figureCV}
\end{figure}

The differences between the spiking behavior of neural networks 
prompted a distinction between sparse and `regular' synchronization, which is often invoked in
theoretical neuroscience~\cite{Wan10,BW12,Brunel2008,Tiesinga2000}.
However, from the viewpoint of the mean field Eqs.~(\ref{eq:fre},\ref{eq:synapses}),
oscillations with high and low $\langle r\rangle/\Omega$ correspond to the same periodic attractor,
i.e., the transition between regular and irregular firing activity is smooth
and does not involve any bifurcation.
We illustrate this for stochastic networks in Figure~\ref{figureCV}.
For both moderate $J=100$ (black symbols), and strong  $J=400$ inhibition,
noise gradually increases the difference between $\Omega$ and $\langle r\rangle$
(see panel a),  as well as the firing irregularity monitored by the CV (see panel b).
For moderate inhibition, neurons remain in a fairly regular regime up to the Hopf bifurcation.
Conversely, for strong inhibition the frequency difference and spike irregularities increase rapidly with $\Gamma$.
If, instead, we fix the total amount of disorder and transition from a heterogeneous
to a stochastic network, then $\Omega$ and $\langle r \rangle$ remain 
constant whereas the CV rapidly changes (see Appendix~\ref{appB1}).

The shape of the Hopf boundaries in Fig.~\ref{figurePD} indicate that
the oscillation region shrinks as coupling increases.
This contrasts with setups using Gaussian noise~\cite{Brunel1999,Tiesinga2000,Brunel2006},
which show a persistence of the oscillatory dynamics for arbitrary strong inhibition and moderate disorder.
In Appendix~\ref{appB2} we show that this is also the case for populations of 
QIF neurons with Gaussian noise or heterogeneity 
(see also~\cite{PP22}). 
Therefore, though Cauchy noise appears to be more disruptive of the network synchronicity,
we find that Gaussian and Cauchy distributions produce the same type of 
dynamical behaviors~\footnote{
There exist no exact low-dimensional reductions 
for populations of QIF neurons with Gaussian disorder,
and the approximated techniques proposed so far rely on weak noise assumptions~\cite{Goldobin2021,RP19,Goldobin2021_2,DiVolo22}.
}.

\section{Conclusions}

Fast neural oscillations with irregular spike discharges at low firing rates 
---the sparse synchronization regime~\cite{Brunel2008}--- are pervasive in brain networks, 
and are successfully reproduced in numerical simulations of  large
spiking neuron networks with delayed inhibition and noise~\cite{Brunel1999,Brunel2000,Brunel2006,Tiesinga2000}.
Yet, the extent to what FRM (that are powerful and broadly used tools for the analysis of 
neuronal dynamics) describe sparse synchrony remains elusive. Moreover, 
sparsely synchronized states are considered to be more robust and at odds with the non-sparse synchronized states emerging in 
deterministic populations of self-sustained oscillators~\cite{Tiesinga2000,Brunel2008,BW12}.

Here we derived an exact FRM ---Eqs.~(\ref{eq:fre},\ref{eq:synapses})--- 
that unambiguously links fast global oscillations 
with the presence of sparse synchronization at the single-cell level.
In addition, we demonstrate that 
precisely the same FRM describes fast oscillations emerging in networks 
of deterministic, self-sustained 
oscillators. Therefore, our results indicate that only the neurophysiological 
mechanisms leading to the emergence of fast neuronal oscillations   
(inhibition, synaptic kinetics and/or delays and sufficient drive to induce spiking) 
 determine the nature of the large-scale dynamics of the network, 
and not the level of regularity of the single neuron spiking activity.
In~\cite{Tiesinga2000} a similar equivalence between 
noise (Gaussian) and heterogeneity (uniform) was numerically observed.
However, that analysis of networks with heterogeneity did not include the 
case of strong coupling, and the main focus was put on stochastic networks.

Altogether, our results reconcile two traditionally confronted views
on the nature of fast neural rhythms (sparse vs. non-sparse
synchronization)~\cite{Tiesinga2000,Brunel2008,BW12}
and upgrades the applicability of exact FRM for QIF
neurons to describe a broad range of biologically realistic neuronal states.
Furthermore, our results can be readily extended to incorporate noise 
in a variety of extensions of the QIF-FRM, such as in
interacting communities of excitatory and 
inhibitory populations~\cite{Montbrio2015}, or in populations with 
conductance-based~\cite{Coombes2019}, or electrical synapses~\cite{Laing2015,Pietras2019,Montbrio2020}. 
\begin{acknowledgements}
The authors thank Jordi Garcia-Ojalvo for helpful discussions. 
PC acknowledges financial support from the European Union’s Horizon 2020 
research and innovation programme under grant agreement No 101017716 (Neurotwin).
EM acknowledges support by the Agencia Estatal de Investigaci\'on 
under the Project No.~PID2019-109918GB-I00.

\end{acknowledgements}

\appendix

\section{Cauchy noise as a limit of a Poisson process}\label{appA}

Here we discuss a different interpretation of the Cauchy noise, which leads to a derivation of the mean-field theory that avoids the use of the Riesz operator.
Let us assume now a more general case of Eq.~\eqref{eq:qif} in which the random variables $\xi_j(t)$ in Eq~\eqref{eq:qif} correspond to a
Poisson shot process with rate $\nu$ and independent random increments given by a probability density $F(u)$.
Then, the macroscopic equation for the time evolution of $P$ is given by the Generalized Fokker-Planck Equation (GFPE)~\cite{Denisov2008,klafter2011,Mendez2014,Tonjes2020}
\begin{equation}
\begin{aligned}\label{eq:gfpe}
	\tau_m &\frac{\partial P}{\partial t}(V,t;\eta) =
	-\frac{\partial}{\partial V}\left\{\left[ V^2+\eta-\tau_m Js(t)\right] P(V,t;\eta)\right\}\\
& +\nu\int_{-\infty}^\infty F(u)P(V-u,t;\eta)\dd u - \nu P(V,t;\eta)\;.
\end{aligned}
\end{equation}
The right-hand-side of this integro-partial differential equation contains three terms.
First, the advection term corresponding to the deterministic flow in Eq.~\eqref{eq:qif}.
Second, a convolution integral accounting for the increase of probability due to the Poisson shot process
with rate $\nu$ and increments $F(\cdot)$. This can be interpreted as a \emph{source} term in a continuity equation.
And third, the loss of probability with rate $\nu$ due to the stochastic dynamics, which can be interpreted as a \emph{sink} term.\\

We consider $F(\cdot)$ a Lorentzian distribution centered at zero and with half-width at half maximum $\Gamma\nu^{-1}$,
\begin{equation}
        F(u):=\frac{1}{\pi}\frac{\Gamma\nu }{(\nu u)^2+\Gamma^2}\;.
\end{equation}
In this case, the limit $\nu \to \infty$ corresponds to $\xi$ being the Cauchy white noise used in the main text.
Indeed, in this limit, the GFPE~\eqref{eq:gfpe} corresponds to the FFPE~\eqref{eq:ffpe},
as we shall prove next.

Let us rewrite the GFPE~\eqref{eq:gfpe} as
\begin{equation}
\begin{aligned}
	\tau_m \frac{\partial P}{\partial t}(V,t;\eta)
	=&-\frac{\partial}{\partial V}\left\{\left[ V^2+\eta+\tau_m Js(t)\right] P(V,t;\eta)\right\}\\
	&+\nu [P\ast (F-\delta)](V) \;.
\end{aligned}
\end{equation}
where $\delta$ is a Dirac delta function and $\ast$ is the convolution operator
\begin{equation}
        [f\ast g](x)=\int_{-\infty}^{\infty} f(y)g(x-y)\dd y\;.
\end{equation}
Then, the Fourier transform of the stochastic term of the GFPE reads
\begin{equation}
        \mathcal{F}\left\{\nu [P\ast (F-\delta)](V) \right\} =\nu \tilde P(k,t) (\tilde F(k)-1).
\end{equation}
where $\tilde P(k,t)=\mathcal{F}\{P(V,t)\}$ and $\tilde F(k)=\mathcal{F}\{F(V)\}$.
Since $F$ is a Lorentzian, $\tilde F(k)=\exp( -i|k|\Gamma \nu^{-1})$.
Therefore
\begin{equation}
        \lim_{\nu\to\infty} \nu \tilde P(k,t)(\tilde F(k)-1)=-\Gamma |k|\tilde P(k,t)\;,
\end{equation}
i.e., the integral term in the GFPE~\eqref{eq:gfpe} corresponds to the Riesz fractional derivative~\eqref{eq:riesz} with $\alpha=1$.\\

To conclude, we show that our results still hold if the limit $\nu\to\infty$ is taken after computing the integral term in the GFPE~\eqref{eq:gfpe}.
Using the ansatz~\eqref{eq:ansatz}, the integral on the right hand side of the GFPE~\eqref{eq:gfpe} corresponds to the convolution of two Lorentzian distributions.
Since the sum of two Cauchy random variables also follows a Cauchy distribution (see, e.g.,~\cite{Johnson1994-it}), such convolution integral is readily solved as,
\begin{equation}
       \int_{-\infty}^\infty F(u)P(V-u,t;\eta)\dd u =
       \frac{1}{\pi}\frac{ x +\Gamma\nu^{-1}}{ (V-y)^2 + (x+\Gamma\nu^{-1})^2 }\;.
\end{equation}
Therefore
\begin{align}\label{eq:noise}
	\nu&\int_{-\infty}^\infty F(u)P(V-u,t;\eta)\dd u  -\nu P(V,t;\eta)\\
       &=\frac{\nu}{\pi}\left[\frac{x+\Gamma\nu^{-1}}{ (V-y)^2 + ( x+\Gamma \nu^{-1})^2 }-\frac{x }{(V-y)^2 + x^2}\right]\\
	&\fletxa \frac{\Gamma}{\pi}\frac{(V-y)^2 - x^2}{[(V-y)^2 + x^2]^2}
\end{align}
which coincides with the Riesz derivative of $P$ given in Eq.~\eqref{eq:equiv} multiplied by the noise coefficient $\Gamma$.

\section{Supplementary numerical results}

\subsection{Combining heterogeneity and noise}\label{appB1}

Figure~\ref{figureCV} of the main text shows how increasing noise in a network of QIF neurons
smoothly transitions the system from
a state in which the neurons fire regularly (CV$\approx 0$) to
a irregular microscopic activity (CV$\approx 1$).
Here we provide further evidence of the equivalence, at the collective level,
of noise and heterogeneity, in spite of clear differences
in the spiking regularity of single neurons.

We performed simulations of $N=8192$ described by Eq.~\eqref{eq:qif}
keeping the level of disorder constant at $\Delta+\Gamma=3.5$,
but varying the ratios of noise and heterogeneity.
In order to do so we introduce a new parameter $p\in[0,1]$ quantifying
the amount of heterogeneity in the network, i.e., $p=\Delta/(\Delta+\Gamma)$.

\begin{figure}[t]
  \centerline{\includegraphics[width=0.45\textwidth]{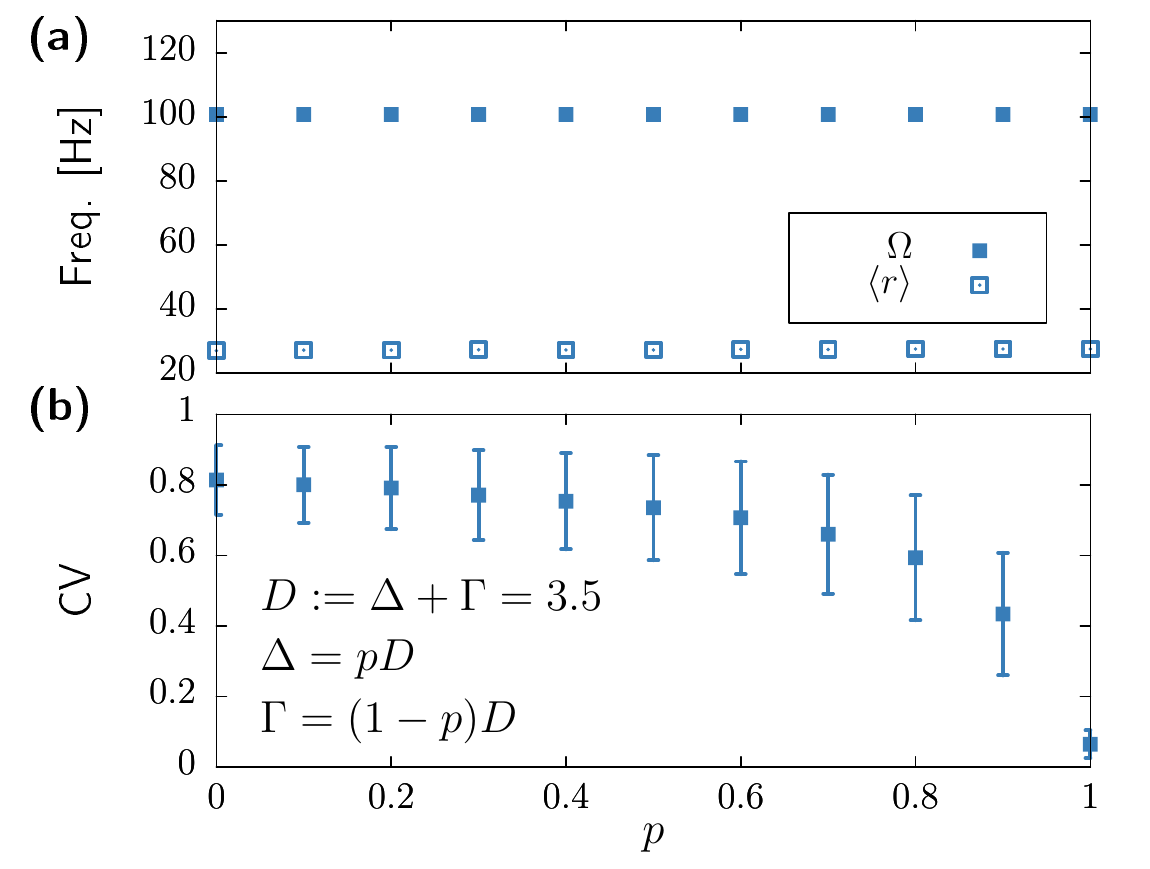}}
        \caption{Oscillation frequency $\Omega$ and time-averaged firing rate $\langle r \rangle$
        of a network QIF neurons with fixed disorder $\Gamma+\Delta=3.5$ and varying the amount of heterogeneity $p$.
        Results obtained integrating Eqs.~(\ref{eq:fre},\ref{eq:synapses}) of the main text for $J=400$ and the rest of the parameters set as in the main paper.
        (b) Average coefficient of variation (CV) of the ISI corresponding to the same simulations.
        Error bars indicate sample standard deviation.
   }
  \label{figureA1}
\end{figure}

The results are depicted in figure~\ref{figureA1}, which shows the collective frequency $\Omega$
and the time-average mean firing rate $\langle r \rangle$ (panel a);
and the average coefficient of variation (CV, panel b).
In order to compute the average CV, neurons with less than 2 spikes have been discarded from the computation
(since at least 2 spikes are needed to obtain at least one ISI).

For $p=0$ we recover the case depicted in column (d) of Figure~\ref{figure1} of the main text, in which neurons
fire irregularly, and thus CV$\approx 1$.
As the amount of heterogeneity increases in the network, the regularity of the firing also
increases, corresponding to a smooth decrease of the CV, which attains CV$\approx 0$
for the full deterministic case ($p=1$). However, both, the macroscopic
oscillatory frequency $\Omega$ and the time-average mean firing rate $\langle r\rangle$
remain unchanged by $p$, as can be inferred from the FRM (Eqs. (\ref{eq:fre},\ref{eq:synapses})).

\subsection{Numerical results with Gaussian heterogeneity and noise}\label{appB2}

The remarkable analytical properties of Cauchy-Lorentz distributions
allow one to derive exact mean-field equations for QIF neurons with such distributions of noise and heterogeneity.
However, in simulations of spiking neurons it is usually more common to use
Gaussian distributions due to their apt statistical properties.
Unfortunately, to date, no exact low-dimensional reduction exists for QIF neurons with Gaussian
heterogeneities or noise, although notable progress has been made in this direction.
Next we discuss the case of Gaussian heterogeneity and Gaussian noise separately.

In~\cite{PP22} the authors proposed exact mean-field equations
for networks of QIF neurons with $q$-Gaussian heterogeneities.
Such distributions are indexed by a parameter $n=1,\dots,\infty$, ($n=(q-1)^{-1}$),
for which $n=1$ corresponds to Lorentzian heterogeneity
and Gaussian heterogeneity is achieved in the limit $n\to\infty$.
The dimensionality of the resulting firing rate equations is $2n$, plus an additional equation
for the synaptic dynamics Eq.~\eqref{eq:synapses}.
Therefore, a network with purely Gaussian heterogeneities remains described by an infinite-dimensional system.

\begin{figure*}[t]
  \centerline{\includegraphics[width=\textwidth]{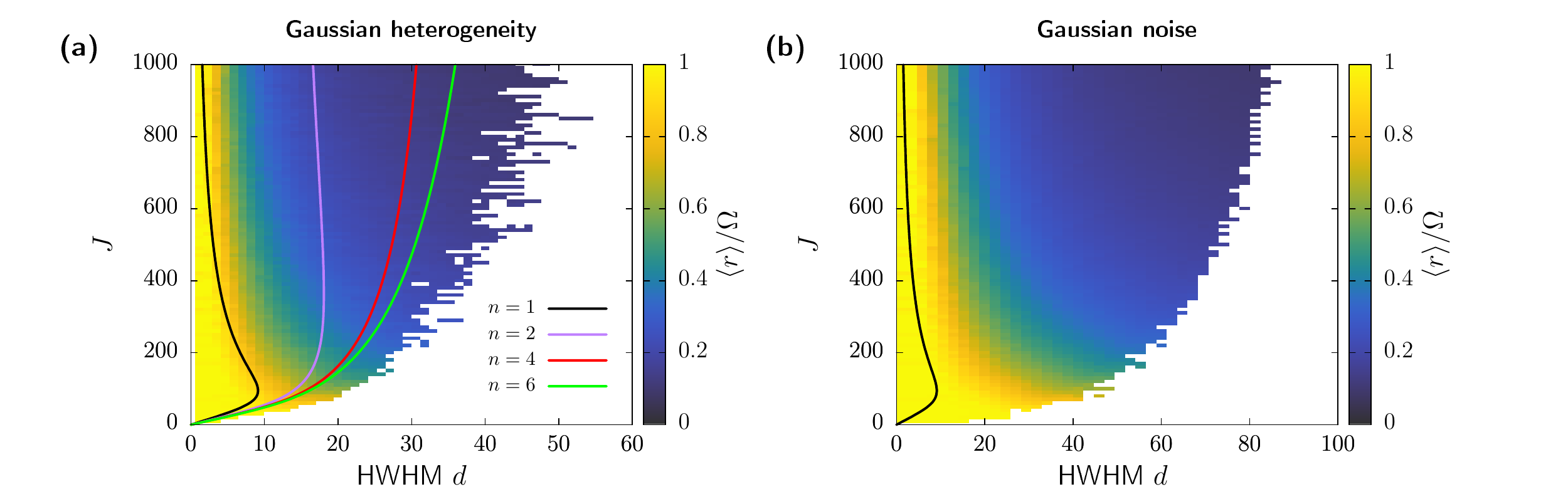}}
        \caption{Phase diagram of the QIF network with Gaussian distributions.
        (a) Lines: Supercritical Hopf bifurcations of the mean-field equations in~\cite{PP22}
	for different values of $n$ (parameter values as in Fig.~\ref{figurePD}).
        The colormap corresponds to the average number of spikes per oscillation cycle for
        a QIF network with Gaussian heterogeneity. It has been numerically obtained through
        numerical integration of the QIF network Eqs. (\ref{eq:fre},\ref{eq:synapses}) of the main paper ($N=8192$).
        (b) Black line as in Fig. 3 of the main paper.
        Colormap corresponds to simulations of QIF neurons ($N=8192$) with Gaussian white
        noise and without heterogeneity (rest of the parameters as in the body of the paper).
        In both panels, a simulation has been considered to be in the oscillatory regime if the standard deviation
        of the synaptic activity $s(t)$ is below $10^{-3}$.
   }
  \label{figureA2}
\end{figure*}

In figure~\ref{figureA2}(a) we show a bifurcation diagram for QIF neurons
with $q$-Gaussian heterogeneites for different values of $n$.
We use the half-width at half-maximum (HWHM) $d$ as control parameter of the heterogeneity.
Notice that $d=\Delta$ for a Cauchy distribution ($n=1$) and $d=\sigma \sqrt{2\ln(2)}$ for a Gaussian
distribution with standard deviation $\sigma$ ($n\to\infty$).
Black line in figure~\ref{figureA2} corresponds to the black continuous line in Fig.~\ref{figurePD}.
As $n$ increases, the region of oscillations widens.
The colormap shows the level of activity in the region of oscillations obtained using
simulations of Eqs.~(\ref{eq:qif},\ref{eq:synapses}), but considering $\eta_j$ to be distributed
as a Gaussian with mean $\overline \eta$ and standard deviation $\sigma$.
In spite of the enlargement of the instability region,
the Hopf bifurcation remains displaying two different cases:
For low coupling $J$ most neurons remain active even at the bifurcation line ($\Omega\approx \langle r\rangle$).
Instead, for high coupling there is a large degree of suppression as $d$ increases.

For Gaussian noise, attempts to derive mean-field equations have been put forward~\cite{Goldobin2021,RP19,Goldobin2021_2,DiVolo22}.
However, these theories build on weak noise approximations ($\sigma \ll 1$)
and are thus unsuitable to analyze networks with large fluctuations.
Figure~\ref{figureA2}(b) displays a numerical bifurcation
diagram of the QIF network (Eqs.~(\ref{eq:qif},\ref{eq:synapses})) with $\xi_i(t)$ being Gaussian white noise with standard deviation $\sigma=d/\sqrt{2\ln(2)}$.
The scenario remains remarkably similar to the case of Gaussian heterogeneity,
with still two qualitatively different transitions towards stationarity as $d$ is increased.
It is worth noting that we did not find cluster states.
This contrasts with the results of~\cite{Brunel2006}, which analyze a similar setup with
other integrate-and-fire models and find cluster instabilities for very low levels of Gaussian noise.
A possible explanation might be the lack of a fixed delay and/or a rise synaptic time in our modelling setup.

\bibliography{references}

\end{document}